\RequirePackage{lineno}
\documentclass[aps,prb,twocolumn,superscriptaddress]{revtex4}

\usepackage{graphicx}
\usepackage{bm}
\usepackage{amsmath}
\usepackage{amssymb}
\usepackage{hyperref}

\begin{document}

\title{Effect of size distribution on the adhesion of nanoscale asperities}

\author{Xiaolong Yang}
\affiliation{Guangxi Key Laboratory for Relativistic Astrophysics, Department of Physics, Guangxi University, Nanning 530004, P. R. China.}
\affiliation{Frontier Institute of Science and Technology, Xi'an Jiaotong University, Xi'an 710049, P. R. China.}

\author{Zhao Wang}
\email{wzzhao@yahoo.fr}
\affiliation{Guangxi Key Laboratory for Relativistic Astrophysics, Department of Physics, Guangxi University, Nanning 530004, P. R. China.}
\affiliation{Frontier Institute of Science and Technology, Xi'an Jiaotong University, Xi'an 710049, P. R. China.}

\begin{abstract}

We study the effect of asperity size on the adhesion properties of metal contact using atomistic simulations. The simulated size effect of individual nanoscale asperityies is applied to macroscopic rough surfaces by introducing a curvature radius distribution to a continuum-mechanics-based contact model. Our results indicate that the contact adhesion can be optimized by changing the curvature radius distribution of the asperity summits. This would open the door
to enhanced metal contact via surface nanostructuring. 

\end{abstract}

\maketitle


Asperities at the atomistic scale are brought into contact when two macroscopic solids touch. Due to surface roughness, the true contact area $A_{c}$ usually only holds a minuscule fraction of the apparent one $A$. The ratio $A_{c}/A$ determines the transfer efficiency of load, current and heat across the interface, and is thus crucial for many technological applications. In the past, a lot of effort has been devoted to studying contact at nanoscale by means of microscopy experiments as well as atomistic simulations. A number of exciting features of nanoscale contacts regarding adhesion,\cite{Luan05,Lu10,Klajn07,Tang02} plasticity,\cite{li02} friction,\cite{Mo09} elasticity\cite{sun14nat,Wang12,Akarapu2011} and strength\cite{Shan2008,kim10prl} have been reported. A major challenge remains, however, to bridge the gap between these nanoscale features and the corresponding characteristics of their macroscopic counterparts. 

Previously established theories based on continuum mechanics\cite{Johnson71,Derjaguin75} make the possibility of studying adhesion properties of microscopic contacts. For instance, Fuller and Tabor\cite{Fuller1975} have generalized the Greenwood-Williamson theory\cite{greenwood66} to adhesive contact by including the Johnson-Kendall-Roberts (JKR) model.\cite{Johnson71} Bush and co-workers have further extended this theory to multi-scale roughness by approximating the summits by random paraboloids of the same principal curvature.\cite{Bush1975} Persson has developed a multi-scale approach considering interactions between asperities.\cite{persson06,Persson2008} Robbins and co-workers have proposed analytical\cite{Pastewka2013} and numerical\cite{Luan05} scaling approaches to study the mechanical response of contacts to external loads, and explained the non-sticky to sticky contact transition.\cite{pastewka14pnas} While these theories are revealing, the assumptions of linear elasticity and uniform curvature radius could be limiting, especially close to the lower wavelength cutoff ($\sim 10$nm) due to diffusion- or dislocation-induced plasticity at this scale.\cite{Luan05,Mo09,Guo2015} It is in this context that the combination of atomistic simulations and continuum contact theory becomes the most valuable in studying nanometer-size effects on solid adhesion.\cite{Luan2005a}


Here we combine large-scale atomistic simulations and classical contact scaling theory by introducing a curvature radius distribution into macroscopic rough surfaces. In our simulations, the atomistic interaction is described by an embedded atom method (EAM) potential with parameterizations from Refs.\onlinecite{Zope03,Mishin99}. The inter-cylinder long-range interactions are described by a Lennard-Jones potential\cite{Pound75} with parameters customized by fitting to the relation between the energy and inter-atomic separation given by the EAM potential in short range interaction. Using the classical parallel simulator LAMMPS,\cite{Plimpton95} we perform molecular mechanics simulations\cite{Boeyens2001,Wang2007,wangpbr2007,Wang09,Wang2010b} to obtain the ground-state ($T \cong 0$K limit) configurations of the contact by minimizing the total potential energy using the conjugated gradient algorithm.


\begin{figure}[htp]
\centerline{\includegraphics[width=9cm]{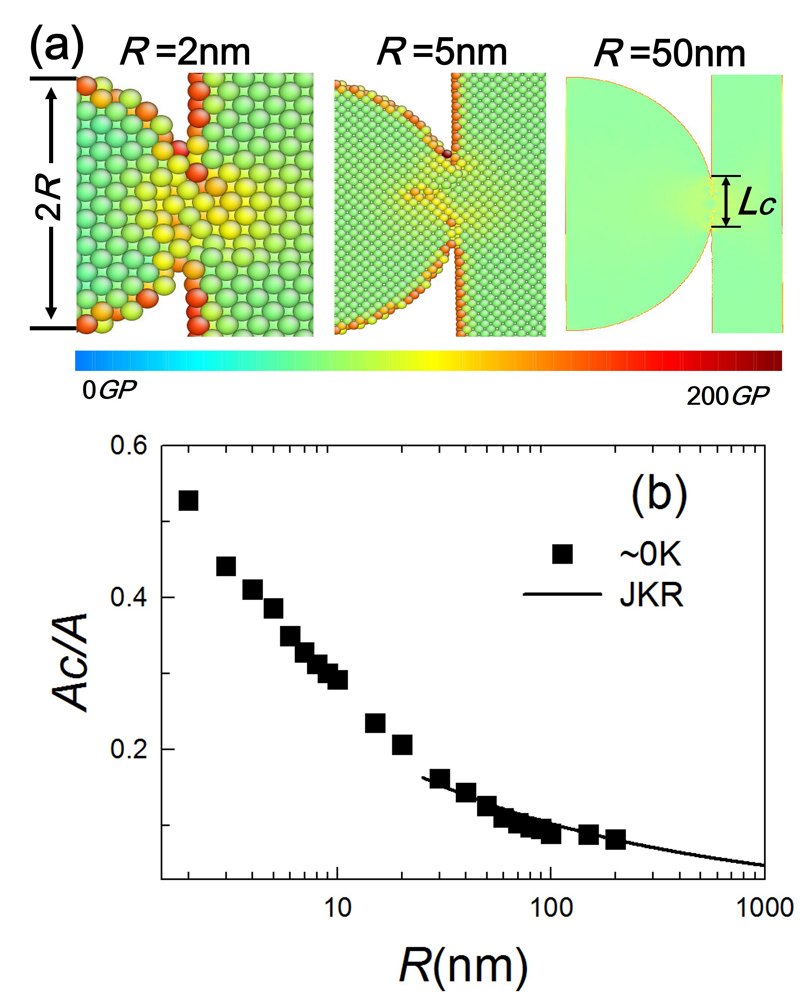}}
\caption{\label{fig:1}
(a) Cross-sectioned snapshots of contacts with different curvature radii $R$. The color scale corresponds to the von Mises stress distribution. (b) Contact area ratio between a flat surface and a curved asperity at $T$ $\cong 0$K as a function of $R$. The symbols stand for simulation results and the curve represents numerical fitting data (Eq.\ref{eq:1}).}
\end{figure}

We start by simulating the adhesive contact between a flat surface and an asperity with a curved tip following a nanowire indentation setup,\cite{Wang12} in which the asperity is spontaneously attached to the flat surface by atomistic interactions at zero external load (so-called spontaneous adhesion) [Fig.\ref{fig:1}(a)]. The effective contact area $A_{c}$ is explicitly computed by defining an inter-cylinder spacing cutoff of $0.286$nm as the equilibrium inter-atomic distance.\cite{Cheng2010} The ratio between the effective contact area $A_{c}$ and the apparent one $A=2Rw$ ($w$ being the width) is computed and shown in Fig.\ref{fig:1}(b) as a function of the curvature radius $R$. We see that $A_{c}/A$ is greatly enhanced by decreasing $R$ of the contacting asperities; this effect becomes most pronounced for tip radii below $10$nm. This enhancement is directly related to the electrostatic nature of the inter-atomic force, since a sharp tip (small $R$) means that a larger fraction of surface atoms are exposed within the attractive interaction distance cutoff, while the inter-atomic forces decrease rapidly with increasing separation distance and vanish after a few nanometers.

For relatively large asperities, the JKR theory\cite{Johnson71} can be used to calculate $A_{c}/A$ taking into account the surface energy $\gamma$ given by the atomistic simulation,

\begin{equation}
\label{eq:1}
\frac{A_{c}}{A} =\left(\frac{6\pi \gamma}{KR} \right)^{\frac{1}{3}},
\end{equation}

\noindent where $K$ is an effective elastic constant $K=4/3\pi (k_{1}+k_{2})$ with $k_i=(1-\nu_i^2)/\pi E_i$, $(i=1,2)$. $k_i$, $\nu_i$ and $E_i$ are the elastic constant, the Poisson ratio and the Young moduli of the contacting bodies, respectively. The JKR model is known to become inaccurate when the contact size falls below ten nanometers due to the surface roughness effect,\cite{Luan05,TABOR1977} Thus, it is used here to extrapolate data for large tips with $R>100$nm, which can hardly be treated by atomistic simulations due to limitations in computational resources.

\begin{figure}[htp]
\centerline{\includegraphics[width=9cm]{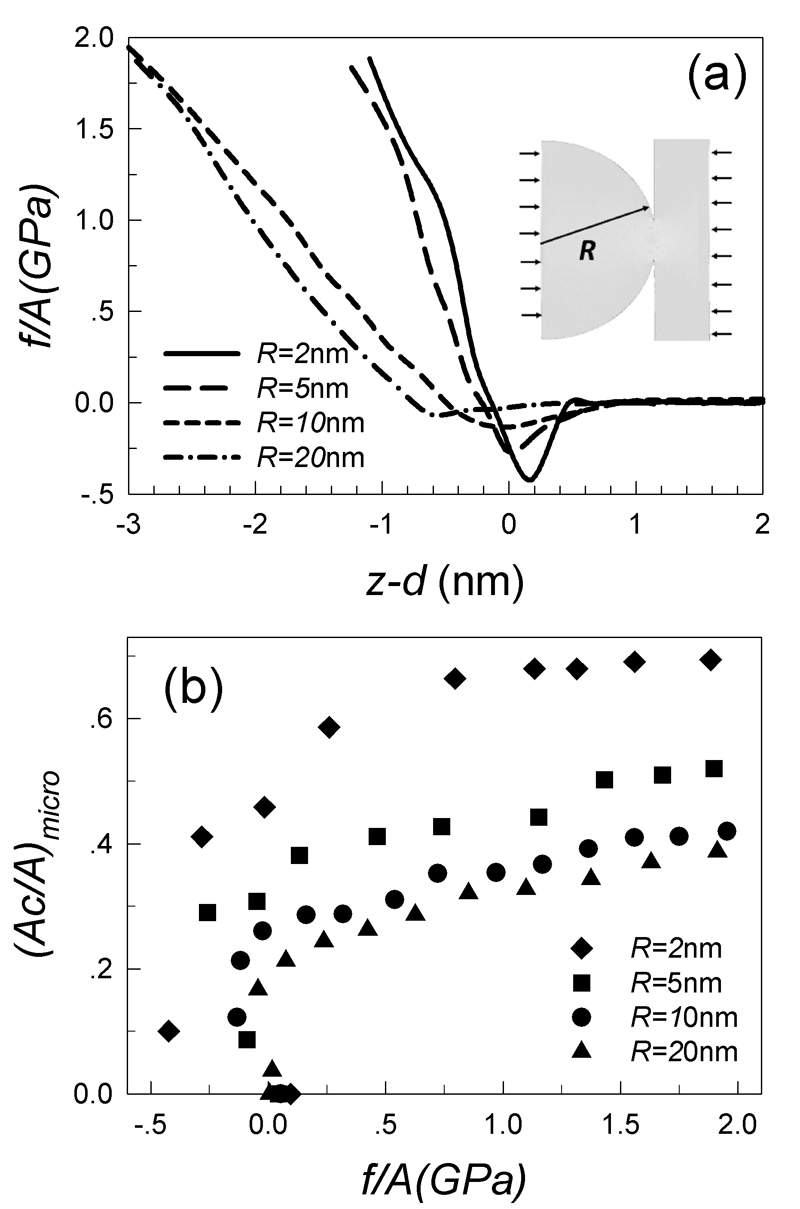}}
\caption{\label{fig:2}
(a) Compression force $f$ per unit area as a function of $z-d$ for adhesive contact between a flat surface and a curved asperity at  $T \cong 0$K. (b) Contact area ratio $(A_{c}/A)_{micro}$ \textit{vs.} force $f$ per unit area in the $T \cong 0$K limit.}
\end{figure}

\begin{figure}[htp]
\centerline{\includegraphics[width=9cm]{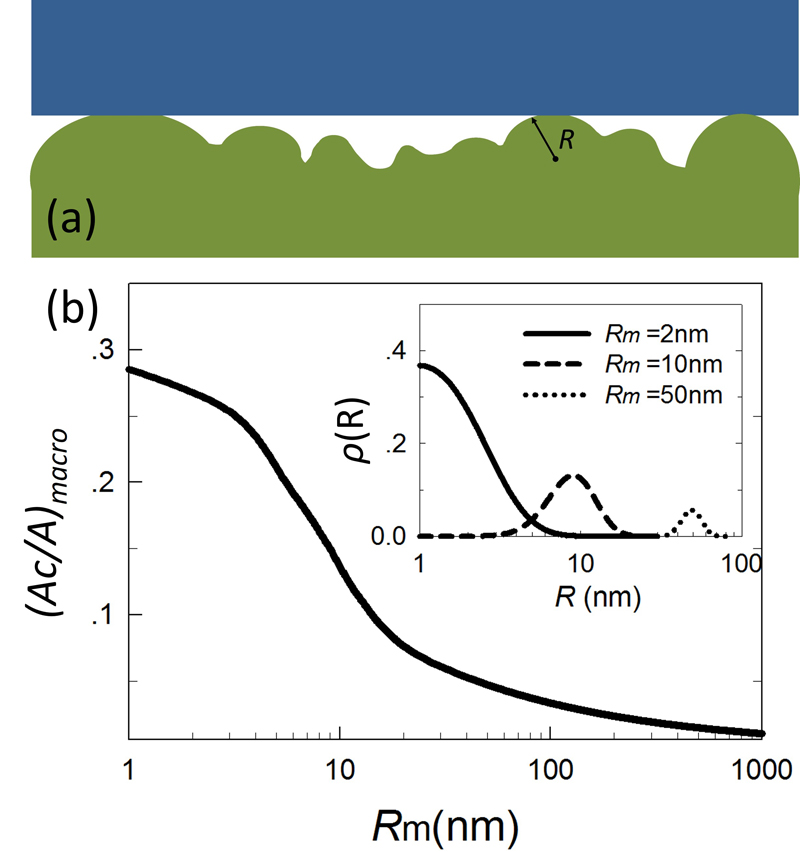}}
\caption{\label{fig:3}
(a) Schematics of contact between a flat and a rough surface. (b) $(A_{c}/A)_{macro}$ \textit{vs.} $R_{m}$. Inset: Probability density $\rho(R)$ for three different mean curvature radii $R_{m}$ with given rate parameters.}
\end{figure}

The above-discussed nanometer size effect is consistent with recent experimentally observed adhesion phenomena of nanomaterials,\cite{Lu10,Tang02,Klajn07} but its applications to real macroscopic contacts however remain limited,\cite{Arzt03,Lee07} due to the length-scale gap along with the well-known surface profile complexity. On the other hand, classical Greenwood-Williamson-based approaches\cite{Fuller1975,greenwood66} deal with uniform surface curvature, while a realistic surface curvature radius distribution is hard to implement directly into more advanced models proposed by Bush\cite{Bush1975} and Persson.\cite{persson06,Persson2008} To this end, we try to bridge the nano and marco scales by approximating a realistic contact surface as a large number of asperities with tip curvature radii distributed in a size range [Fig.\ref{fig:2}(a)]. The contact area ratio of two macroscopic bodies $(A_{c}/A)_{macro}$ with their reference planes separated by a distance $d$ can be written as a collection of those of individual asperities $(A_{c}/A)_{micro}$.

\begin{equation}
\label{eq:2}
\left( \frac{A_{c}}{A}\right) _{macro} = \int\nolimits_{0}^{\infty} \int\nolimits_{d}^{\infty} \rho(R)\phi(z) \left( \frac{A_{c}}{A}\right) _{micro} dz dR,
\end{equation}

\noindent where $\rho(R)$ is a probability density function of curvature radius, and $\phi(z)$ is a curvature height distribution function defined in the Greenwood-Williamson theory. Here the distribution function of curvature radius $\rho(R)$ is assumed to be a Gamma probability density function\cite{JAMBUNATHAN1954} [inset of Fig.\ref{fig:3}(b)], which is chosen because it deals with positive variables and can describe both exponential and Gaussian distributions as particular cases, 

\begin{equation}
\label{eq:3}
\rho(R) =\frac{R^{\alpha-1}e^{-R/\beta}}{\Gamma(\alpha)\beta^{\alpha}},
\end{equation}

\noindent where $\alpha$ and $\beta$ are the shape and scale parameters, respectively, and

\begin{equation}
\label{eq:4}
\Gamma(\alpha)=\int\nolimits_0^{\infty}t^{\alpha-1}e^{-t}dt,
\end{equation}

\noindent where $t$ is a integral variable. The mean curvature radius $R_{m}$ can be calculated as

\begin{equation}
\label{eq:5}
R_{m} = \alpha \beta,
\end{equation}

\noindent with the standard deviation $\sigma$ written as

\begin{equation}
\label{eq:6}
\sigma=\sqrt{\alpha}\beta.
\end{equation}

In the Greenwood-Williamson theory, a random series of asperities with height $z$ is usually represented by a Gaussian distribution, 

\begin{equation}
\label{eq:7}
\phi (z) =\frac{1}{(2\pi)^{\frac{1}{2}}} exp\left(-\frac{z^2}{2\eta ^2}\right), 
\end{equation}

\noindent where $\phi (z)$ represents the probability density that a rough surface has asperities with height around $z$, calculated with respect to the reference plane defined by the mean height, and $\eta$ is the standard deviation of $\phi (z)$. For two contacting rough surfaces with their reference planes separated by a distance $d$, asperities with $z>d$ are assumed to be in contact. In such context, the probability for any asperity in the rough surface to make contact can be written as follows, 

\begin{equation}
\label{eq:8}
prob(z>d) =\int\nolimits_{d}^{\infty}\phi(z)dz.
\end{equation}  

The computation of adhesion area requires knowledge of the compressive force $f$ as a function of $z$. $f(z)$ is directly computed from our simulations for small asperities with $R \leq 100$nm, as shown in Fig.\ref{fig:2}(a). Fig.\ref{fig:2}(b) shows the change of contact area ratio during compressive loading. To expand the data range to large asperities with $R > 100$nm, JKR fitting gives 

\begin{equation}
\label{eq:9}
z-d = \frac{w(K \frac{A_{c}^3}{w^3}+2fR)}{3RKA_{c}} ,
\end{equation}

\noindent here 

\begin{equation}
\label{eq:10}
\left(\frac{A_c}{A}\right)_{micro}=\left(\frac{6\pi \gamma}{KR} \right)^{1/3}\left(\frac{1+\sqrt{1-f/f_{c}}}{2} \right)^{2/3}, 
\end{equation}

\noindent where $f_{c}=-\frac{3}{2}\pi\gamma R$ is the pull-off force with which the surfaces get to be separated.\cite{Carpick1999} 

The total contact force $F$ is computed as the sum of the forces exerted by all asperities with $z>d$ on one of the contacting bodies, 

\begin{equation}
\label{eq:11}
F = \int\nolimits_{0}^{\infty}\int\nolimits_{d}^{\infty} \rho(R)\phi(z)f(R,z) dz dR.
\end{equation}

For spontaneous adhesion, $F$ must be zero. This constraint corresponds a critical spacing $d_{0}$, which leads to 

\begin{equation}
\label{eq:12}
\int\nolimits_{0}^{\infty} \int\nolimits_{d_{0}}^{\infty} \rho(R)\phi(z)f(R,z) dz dR= 0.
\end{equation}
 
Finally, the contact area ratio of two macroscopic bodies with well-defined surface roughness $\rho(R)$ and $\phi(z)$ in the spontaneous adhesion case (zero external load) can be calculated as  

\begin{equation}
\label{eq:13}
\left(\frac{A_{c}}{A}\right) _{macro} = \int\nolimits_{0}^{\infty} \int\nolimits_{d_{0}}^{\infty} \rho(R)\phi(z) \left( \frac{A_{c}}{A}\right) _{micro} dz dR.
\end{equation}

Keeping the scale parameter $\beta$ constant and changing the shape parameter $\alpha$ in the curvature radius distribution $\rho(R)$, one can calculate the effective contact area ratio using Eq.\ref{eq:13}, an example of which is given in Fig.\ref{fig:3}(b). We see that a macroscopic contact can still be remarkably enhanced when the average tip radius $R_{m}$ is lowered below to $10$nm. For instance, we find an increase of one order of magnitude at $R_{m} \approx 1.0$nm. Comparing to theories assuming a uniform curvature, we note that the macroscopic contact area would behave differently under a change of the height distribution $\phi (z)$ by tuning its standard deviation $\eta$ in Eq.\ref{eq:7}.      

\begin{figure}[htp]
\centerline{\includegraphics[width=9cm]{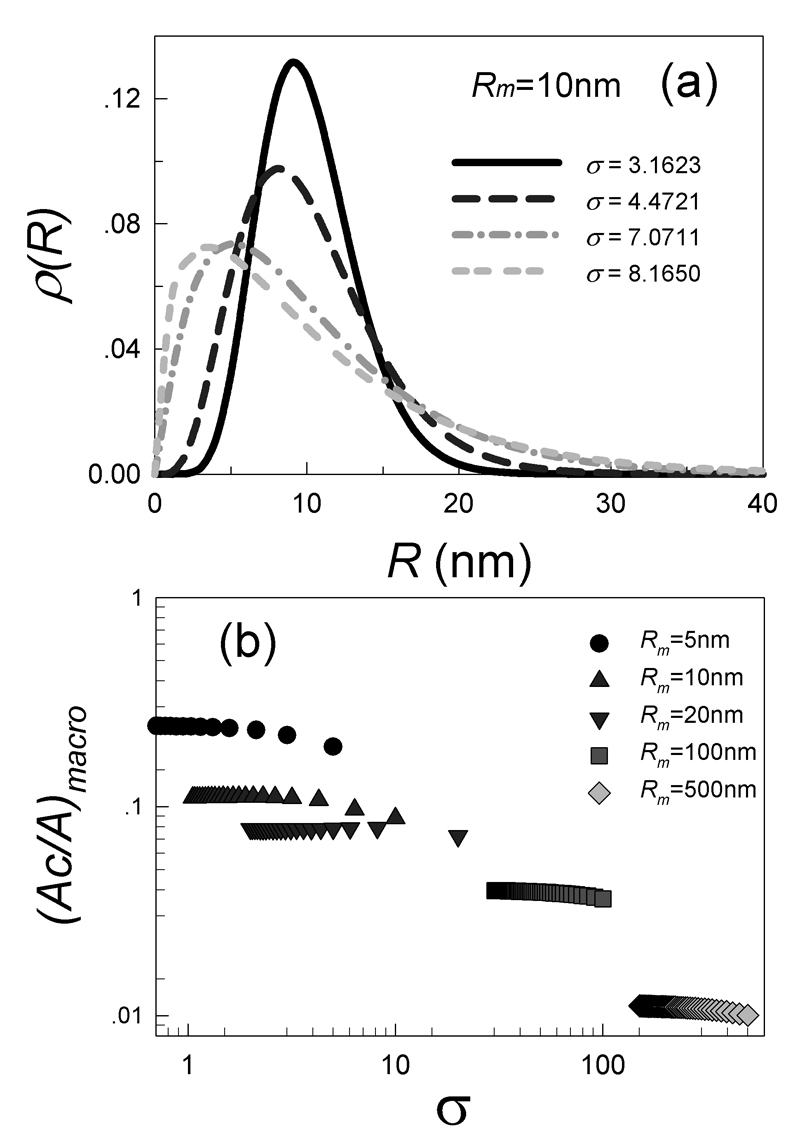}}
\caption{\label{fig:4}
(a) Probability density $\rho(R)$ for different standard deviation $\sigma$ of the gamma distribution at a mean curvature radius $R_{m}=10nm$. (b) $(A_{c}/A)_{macro}$ \textit{vs.} $\sigma$ for different $R_{m}$ with a given standard deviation of the height distribution $\eta=1.0$.}
\end{figure}

Fig.\ref{fig:4}(b) shows the contact area ratio as a function of the standard deviation $\sigma$ of the curvature radius distribution. Generally the larger $\sigma$, the broader the distribution data range, as shown in Fig.\ref{fig:4}(a). We see that contact area ratio decreases when $\sigma$ increases for a variety of given mean curvature radii $R_{m}$. This is because a broader distribution increases the contact probability between large asperities and the substrate, which correspond to smaller values of $(A_{c}/A)_{micro}$. Besides, it can be found that $(A_{c}/A)_{macro}$ is largely weakened by considering the surface roughness due to the small contact probability for asperity summits.

\begin{figure}[htp]
	\centerline{\includegraphics[width=9cm]{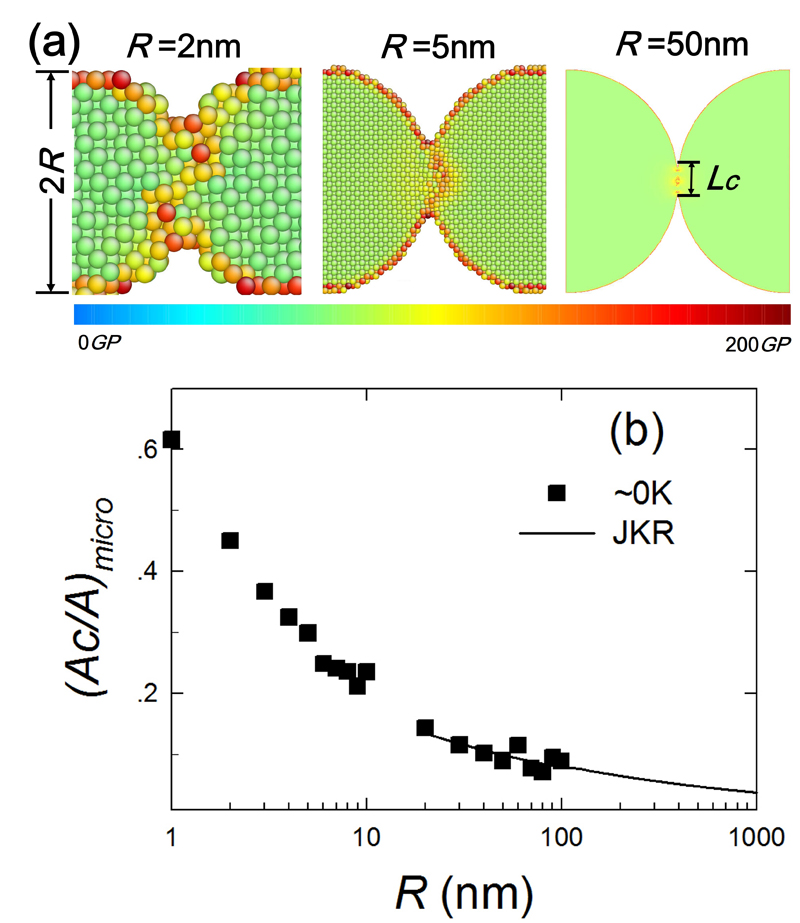}}
	\caption{\label{fig:5}
(a) Cross-sectioned snapshots of contacts with different curvature radii $R$. The color scale corresponds to the von Mises stress distribution. (b) Contact area ratio $(A_{c}/A)_{micro}$ at the ground-state ($T \cong 0$K limit) as a function of $R$. The symbols show simulation results and the curve stands for numerical fitting by Eq.\ref{eq:14}.}
\end{figure}

The above results hold for the relatively simple case of the contact between a flat and a rough surface. The contact between two rough surfaces is however more realistic, as the spontaneous adhesion shown in Fig.\ref{fig:5}(a) for $R=R_{1}=R_{2}$. In Fig.\ref{fig:5}(b), we still see that $(A_{c}/A)_{micro}$ is enhanced when the curvature radius decreases to nanometer scale. The JKR model\cite{Johnson71} is used to provide a rough estimate to the displacive contact area for large tips with $R>100$nm fitting to simulation data, given by

\begin{equation}
	\label{eq:14}
	\frac{A_{c}}{A} =\left[\frac{6\pi \gamma (R_{1}+R_{2})}{KR_{1}R_{2}} \right]^{\frac{1}{3}}.
\end{equation}

\begin{figure}[htp]
	\centerline{\includegraphics[width=9cm]{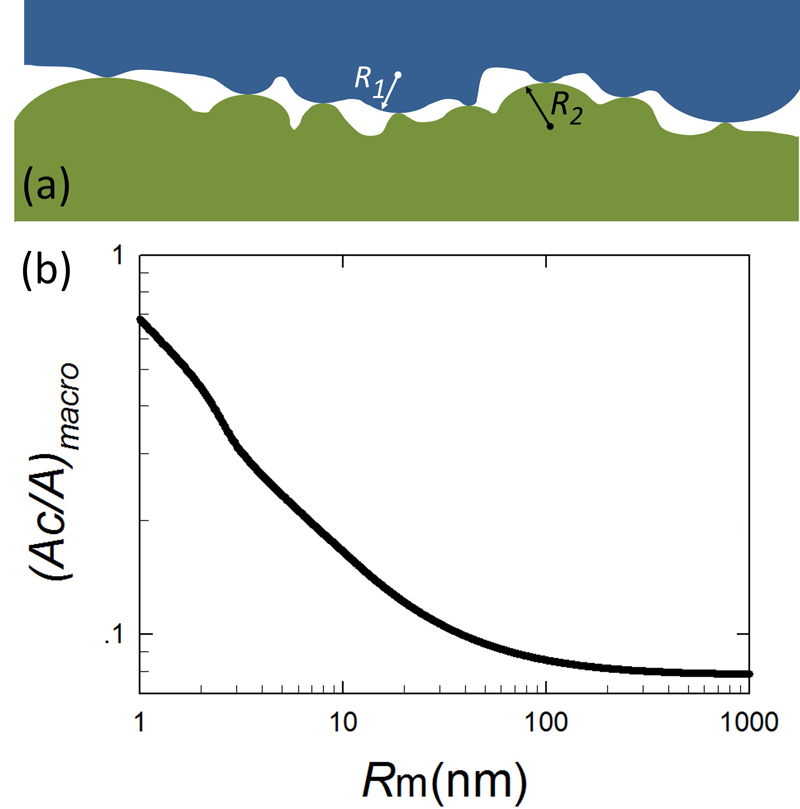}}
	\caption{\label{fig:6}
(a) Schematics of a particular case of contact between rough surfaces with asperities summits in curvature radius distribution $\rho(R_{1})$ and $\rho(R_{2})$. (b) $(A_{c}/A)_{macro}$ \textit{vs.} $R_{m}$ for $\rho(R_{1}) \equiv \rho(R_{2})$.}
\end{figure}

For contact between randomly rough surfaces, it is hard to determine the exact contact probability of each pair of asperities because of the surface profile complexity. Here we consider a highly simplified case in which all asperities at the atomistic scale are in contact at zero external load. Fig.\ref{fig:6}(a) illustrates the schematics of this particular case of the general contact with two curvature radius distributions $\rho(R_{1})$ and $\rho(R_{2})$ introduced. Following the idea of Eq.\ref{eq:13}, we write the macroscopic contact area ratio as

\begin{equation}
	\label{eq:15}
	\left( \frac{A_{c}}{A}\right)_{macro}=\int\nolimits_{0}^{\infty} \int\nolimits_{0}^{\infty} \rho(R_{1}) \rho(R_{2})\left( \frac{A_{c}}{A}\right)_{micro} dR_{1}dR_{2}.
\end{equation}

\begin{figure}[htp]
	\centerline{\includegraphics[width=9cm]{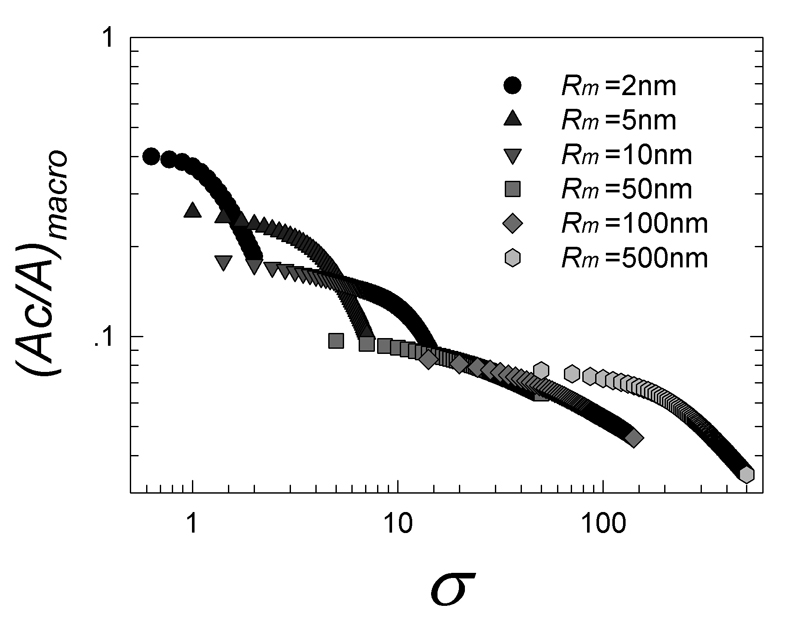}}
	\caption{\label{fig:7}
	 (a) $(A_{c}/A)_{macro}$ between randomly rough surfaces with same the $R$ distribution \textit{vs.} the standard deviation $\sigma$ (Eq.\ref{eq:6}) for different $R_{m}$.}
\end{figure}

By keeping the scale parameter $\beta$ constant and tuning the shape parameter $\alpha$ in the $R$ distribution (Eq.\ref{eq:5}), we obtain the contact area ratio shown in Fig.\ref{fig:6}(b). It shows that the macroscopic contact area rapidly increases when the average tip radius $R_{m}$ reaches below the ten-nanometer scale, in particlar, $(A_{c}/A)_{macro}$ increases by about one order of magnitude at $R_{m} \approx 1.0$nm. 

Fig.\ref{fig:7} shows $(A_{c}/A)_{macro}$ as a function of the standard deviation $\sigma$. We see that the contact area ratio decreases when $\sigma$ increases for a given mean curvature radius $R_{m}$. More importantly, it is very sensitive to the change of the distribution shape, as the contact probability between large asperities increases when the distribution of $R$ is made broader.

\begin{figure}[htp]
	\centerline{\includegraphics[width=9cm]{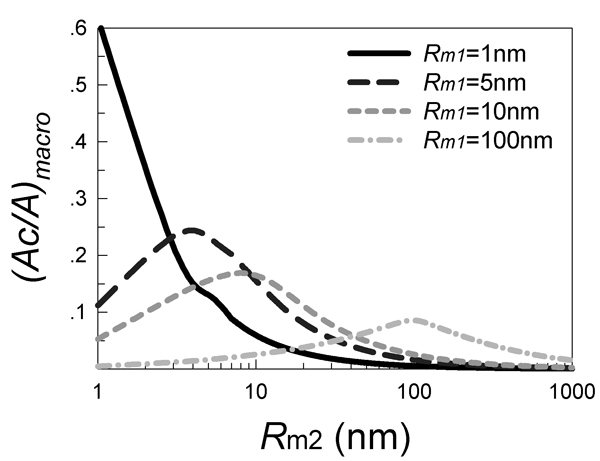}}
	\caption{\label{fig:8}
Contact area ratio $(A_{c}/A)_{macro}$ as a function of $R$ for two adjoining surfaces with different $R$ distributions $\rho(R_{1}) \neq \rho(R_{2})$. The rate parameter $\beta$ is fixed to $1.0$.}
\end{figure}

To make these results more applicable, we shall consider the case of a contact with $\rho(R_{1}) \neq \rho(R_{2})$. Fig.\ref{fig:8} shows that, for a surface with a given distribution $\rho (R_{1})$, there exists an optimal curvature radius distribution of the adjoining surface $ \rho_{opt}(R_{2})$ which maximizes the contact area ratio. We see that $\rho_{opt}(R_{2}) \approx \rho_(R_{1})$ for a $R$ distribution with its mean $R_{m}$ larger than $10$nm, while, the mean of $\rho_{opt}(R_{2})$ is slightly smaller than $R_{m1}$ for $R_{m1}>10nm$ due to the unavoidable surface atom diffusion driven by the very high surface energy of extremely small tips.\cite{Guo2015,Tong2003,Surrey2012,jiang04,Sorensen1996}

Bridging the gap between the unique nanoscale contact features and electrical, thermal and mechanical properties of macroscopic interfaces\cite{greenwood66,persson06,pastewka14pnas,Akarapu2011} requires accurate information about size-dependent plasticity. Strong nanometer-size effects on adhesion make a combination of atomistic simulations and continuum contact theory a powerful tool for bridging the gap between microscopic contacts and their macroscopic counterparts. Our results suggest that the contact adhesion can be optimized by changing the asperity curvature radius distribution of the rough surfaces, and indicate that surface nanostructuring is promising for enhancing solid adhesion. Such an approach could also be applied to material cold welding\cite{Wagle2015,ferguson91science,Lu10} and self assembly,\cite{Klajn07} and may be extended to the modeling of electrical, thermal and mechanical properties of macroscopic solid interfaces.

\section*{Acknowledgments}
We thank Ju Li at MIT for helpful discussions, and Jesus Carrete for proof-reading. This work is supported by the National Natural Science Foundation of China under Grant No. 51571007, the Guangxi Science Foundation (2013GXNSFFA019001), the Guangxi Key Laboratory Foundation (15-140-54), and the Scientific Research Foundation of GuangXi University (Grant No. XTZ160532).


\end{document}